\documentclass[letterpaper]{article} % DO NOT CHANGE THIS
\usepackage{aaai2026}  % DO NOT CHANGE THIS
\usepackage{times}  % DO NOT CHANGE THIS
\usepackage{helvet}  % DO NOT CHANGE THIS
\usepackage{courier}  % DO NOT CHANGE THIS
\usepackage[hyphens]{url}  % DO NOT CHANGE THIS
\usepackage{graphicx} % DO NOT CHANGE THIS
\usepackage{natbib}  % DO NOT CHANGE THIS AND DO NOT ADD ANY OPTIONS TO IT
\usepackage{caption} % DO NOT CHANGE THIS AND DO NOT ADD ANY OPTIONS TO IT
\usepackage{tabularx}
\usepackage{booktabs}
\nocopyright
\usepackage{algorithm}
\usepackage{algorithmic}

\usepackage{newfloat}
\usepackage{listings}
\DeclareCaptionStyle{ruled}{labelfont=normalfont,labelsep=colon,strut=off} % DO NOT CHANGE THIS
\floatstyle{ruled}
\newfloat{listing}{tb}{lst}{}
\floatname{listing}{Listing}
\usepackage{amsmath}
\usepackage{amsfonts}

\title{Sequence Aware SAC Control for Engine Fuel Consumption Optimization in Electrified Powertrain}
\author{
   Wafeeq Jaleel\equalcontrib,
   Md Ragib Rownak,
   Athar Hanif,
   Sidra Ghayour Bhatti,
   Qadeer Ahmed
   }
\affiliations{
   Center for Automotive Research, The Ohio State University\\
    930 Kinnear Road, Columbus, 43212, OH, United States\\
   jaleel.5@osu.edu

}

\begin{document}

\maketitle

\begin{abstract}
As hybrid electric vehicles (HEVs) gain traction in heavy-duty trucks, adaptive and efficient energy management is critical on reducing fuel consumption while maintaining battery charge for long operation times. We present a new reinforcement learning (RL) framework based on the Soft Actor-Critic (SAC) algorithm to optimize engine control in series HEVs. We reformulate the control task as a sequential decision-making problem and enhance SAC by incorporating Gated Recurrent Units (GRUs) and Decision Transformers (DTs) into both actor and critic networks to capture temporal dependencies and improve planning over time. To evaluate robustness and generalization, we train the models under diverse initial battery states, drive cycle durations, power demands, and input sequence lengths. Experiments show that the SAC agent with a DT-based actor and GRU-based critic was within 1.8\% of Dynamic Programming (DP) in fuel savings on the Highway Fuel Economy Test (HFET) cycle, while the SAC agent with GRUs in both actor and critic networks, and FFN actor-critic agent were within 3.16\% and 3.43\%, respectively. On unseen drive cycles (US06 and Heavy Heavy-Duty Diesel Truck (HHDDT) cruise segment), generalized sequence-aware agents consistently outperformed feedforward network (FFN)-based agents, highlighting their adaptability and robustness in real-world settings. 
\end{abstract}


\section{Introduction}
Heavy-duty vehicles are major contributors to fuel consumption and greenhouse gas emissions \cite{muratori2023fuelcont}, and with tightening regulations \cite{DOE-EPA_2025_MHDV}, Series Hybrid Electric Vehicles (SHEVs) offer a better alternative. However, achieving optimal energy distribution between the engine and battery under uncertain driving conditions in real time is still challenging. Early methods used rule-based strategies, later improved by fuzzy-logic control to handle system uncertainties \cite{lee1998fuzzy}. Dynamic Programming (DP) is used for global optimal solutions \cite{Lin2003DP}, but its high computational cost limits any real-time application. The Equivalent Consumption Minimization Strategy (ECMS) \cite{Sciarretta2004ECMS} minimizes fuel consumption by adjusting energy distribution between engine and battery, while Adaptive ECMS \cite{onori2010adaptiveECMS} improved its adaptability. Model Predictive Control (MPC) \cite{sampathnarayanan2009MPC} optimized energy use considering constraints, and Pontryagin’s Minimum Principle (PMP) \cite{kim2010PMP} offered a faster solution, but are still complex for real-time use.\\
\indent Recent advances in computational power and learning algorithms have made data-driven energy management approaches feasible in HEV control \cite{hu2019RLinHEVintro}. Specifically, Reinforcement learning (RL) algorithms such as DDPG, TD3, and SAC have shown better performance, with DDPG improving efficiency but struggling in stability \cite{guo2020DDPG, yao2023DDPG}, and TD3 improving stability by reducing overestimation bias \cite{yao2022TD3HEV, zhou2021TD3HEV}. SAC outperforms both by maximizing reward and entropy (randomness), making it possible to get near-optimal solutions\cite{sun2022SAC, rolando2024SAC, li2022SAC}. \\
\indent However, RL-based HEV controllers typically use feedforward neural networks that ignore temporal dependencies commonly in driving patterns. Temporal sequence modeling within RL, using Recurrent Neural Network (RNN) \cite{liu2024SACRNN}, \cite{ni2021recurrentRL}, and transformer-based models \cite{ludolfinger2023transformer}, \cite{tian2025Transformer}, has been shown to improve RL performance in sequential tasks. Despite success in other domains, these methods have not yet been applied to energy management in powertrain systems.

In this paper, we focus on SAC, as it learns directly from interaction data, handles stochastic environments, and scales well to continuous, high-dimensional control compared to conventional methods such as DP or MPC. We study how sequence-aware architectures, specifically Gated Recurrent Units (GRU) and Decision Transformers (DT), can improve SAC for SHEV energy management. The main contributions are summarized as follows:
\begin{itemize}
\item \textbf{SAC-GRU:} SAC is extended with GRU-based actor and critic networks to capture short- and mid-term temporal patterns using memory components.
\item \textbf{SAC-DT:} Decision Transformers are integrated into SAC for online use, enabling return-conditioned trajectory modeling with causal attention for long-horizon control.

\item \textbf{Ablation Study:} We evaluate the impact of different network architectures, input context lengths, varying battery SOC, drive cycle duration, and power requirements to isolate the effects of sequence-aware SAC.

\item \textbf{Validation:} Trained agents are tested on unseen drive cycles using a high-fidelity MATLAB/Simulink SHEV model, demonstrating improved generalization and fuel efficiency compared to DP baselines.
\end{itemize}

\section{Background}
In this section, we present the specifications of the SHEV and provide a theoretical overview of SAC, GRU, and DT.
\subsection{Series HEV Architecture}
\begin{table}[h]

    \centering
    \resizebox{\columnwidth}{!}{%
    \begin{tabular}{|l|l|}
    \hline
    \multicolumn{2}{|c|}{\textbf{Vehicle}}                                 \\ \hline
    Curb Weight                      & 36287kg\\ \hline
    Wheel Radius & 0.507m\\ \hline
    Frontal Area& 8.48$m^2$\\ \hline
    \multicolumn{2}{|c|}{\textbf{Engine}}                                  \\ \hline
    Max Power                        & 270 kW @ 2300 rpm\\ \hline
    Max Torque                       & 1500 Nm @ 1120-1480 rpm\\\hline
 \multicolumn{2}{|c|}{\textbf{Generator}}\\\hline
 Max Power/ Max Torque            &240 kW @ 2200 rpm / 1410Nm @ 1300rpm\\\hline
 Max Speed                        &2517 rpm\\\hline
    \multicolumn{2}{|c|}{\textbf{Electric Machine}}                        \\ \hline
    Max Power/ Max Torque            & 400kW @ 2000 rpm / 3500Nm @ 1100rpm\\ \hline
    Max Speed                        & 3900rpm\\ \hline
    \multicolumn{2}{|c|}{\textbf{Battery}}                    \\ \hline
    Type                             & NMC\\ \hline
    Rated Voltage (cell)& 3.63 V\\ \hline
    Capacity (cell)& 323.94kWh/ 4.85 Ah\\ \hline
 Cells in S/P&160/115\\\hline
    \end{tabular}
    }
    \caption{Specifications of the Vehicle}
    \label{tab:vehicle_specifications}
\end{table}
Table \ref{tab:vehicle_specifications} provides the specifications for the SHEV used in the simulations, and Figure \ref{fig:SASAC_ARCH} shows the SHEV architecture attached to the overall SAC diagram. The components are modeled as described in HEV: Energy Management strategies \cite{onori2016hybrid}. Equation \ref{eq:HEV constraint} models the HEV’s constraints, where x is the name of the map-based components (Engine, Generator, Electric Machine). The battery's State of Charge (SOC) must remain between 0\% and 100\%, and the energy supplied by the engine and battery must equal the energy requested by the driver, including the energy losses.

\begin{equation}
\label{eq:HEV constraint}
\begin{aligned}
\text{Torque}_{x,\text{min}} &\leq \text{Torque}_x(t) \leq \text{Torque}_{x,\text{max}} \\
\text{Speed}_{x,\text{min}} &\leq \text{Speed}_x(t) \leq \text{Speed}_{x,\text{max}} \\
\text{SOC}_{\text{min}} &\leq \text{SOC}(t) \leq \text{SOC}_{\text{max}} \\
\text{Total Power Demand}(t) &= \text{Power Provided}(t)
\end{aligned}
\end{equation}

\subsection{Soft Actor-Critic (SAC)}

Soft Actor-Critic (SAC) is an off-policy actor-critic algorithm that optimizes a stochastic policy in continuous action spaces using the maximum entropy framework \cite{Haarnoja2018soft}. Unlike standard RL algorithms that maximize expected return, SAC maximizes a trade-off between return and policy entropy, encouraging exploration:

\begin{equation}
\label{eq:SAC_Eq}
J(\pi) = \sum_{t=0}^{\infty} E_{(s_t, a_t) \sim \rho_{\pi}} \left[ r(s_t, a_t) + \alpha \mathcal{H}(\pi(\cdot|s_t)) \right]
\end{equation}

where $\alpha$ is a temperature parameter controlling the exploration-exploitation trade-off. The algorithm maintains:

\begin{itemize}
    \item An actor network $\pi_\theta(a|s)$ that samples actions.
    \item Two critic networks $Q_{\phi_1}(s,a)$ and $Q_{\phi_2}(s,a)$ that estimate state-action values.
    \item Two Q target network for stability.
\end{itemize}

\subsection{Gated Recurrent Units (GRU)} 
GRUs are a type of recurrent neural network designed to capture sequential patterns with a memory mechanism and update hidden states over time \cite{Cho2014GRU}. As seen in Figure \ref{fig:GRU_DT_Architectures} (a), the input consists of a sequence of states $[s_{t-k}, ..., s_t]$, and the update gate $z_t$ determines how much past information to retain, while the reset gate $r_t$ decides how much to forget. The candidate state $\tilde{h}_t$ represents the current input. These components are combined to produce the final hidden state $h_t$,  as a context vector summarizing relevant past information. $h_t$ then passed through a fully connected layer (mean, std) to get the final action output $a_t$. While the GRU-based critic would take both state and action as input to evaluate the quality of actions over time. 

\subsection{Decision Transformers (DT)}

\begin{figure}[htbp]
    \centering
    \begin{minipage}{0.3\textwidth}
        \centering
        \includegraphics[width=1\linewidth]{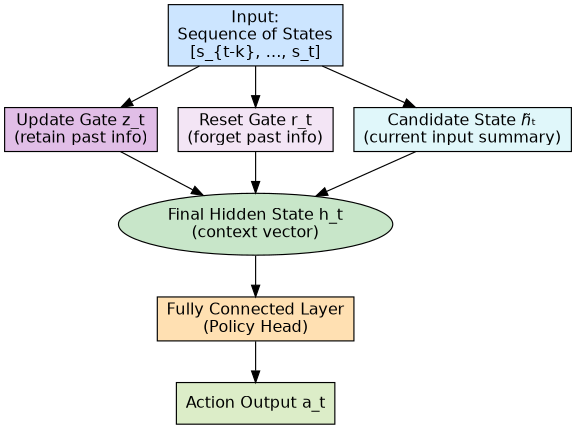}
        % \caption{GRU Architecture}
        \label{fig:GRU_ARCH}
    \end{minipage}%
    \begin{minipage}{0.2\textwidth}
        \centering
        \includegraphics[width=0.6\linewidth]{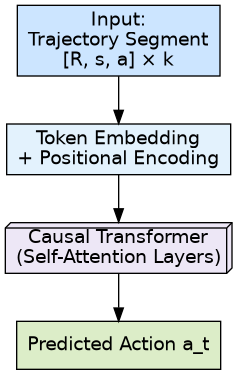}
        % \caption{DT Architecture}
        \label{fig:DT_Architecture}
    \end{minipage}
    \caption{(a) GRU Architecture and (b) DT Architecture}
    \label{fig:GRU_DT_Architectures}
\end{figure}

DT reformulates RL as a sequence modeling problem \cite{Chen2021decision}. Instead of learning value functions directly, it uses a causal attention mechanism to focus on relevant parts of the past sequences of desired return-to-go, states, and actions, which would maximize future returns to predict the next action. This approach helps in a complex environment by capturing long-term dependencies. Figure \ref{fig:GRU_DT_Architectures} (b) shows the model architecture, where the input trajectory segment \([R, s, a] \times k\) goes through token embedding and positional encoding before passing through a causal transformer to predict the next action \(a_t\). While it is designed for offline RL applications, incorporating it within SAC helps it transform into an online RL method. 

\begin{figure*}[htbp]
    \centering
    \includegraphics[width=1\linewidth]{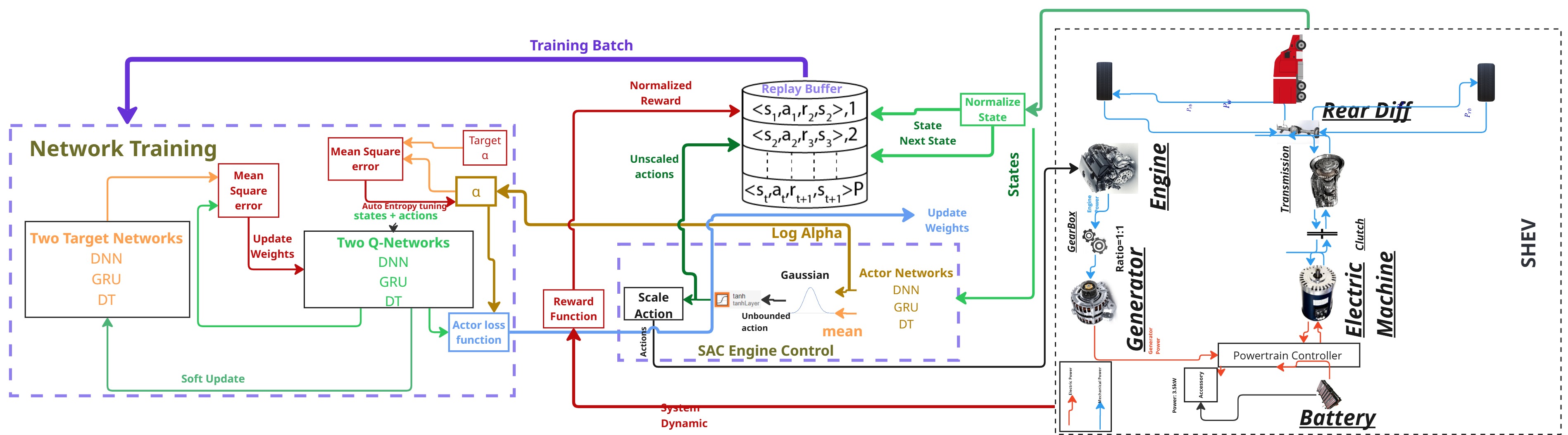}
    \caption{SA-SAC Arhchitecture}
    \label{fig:SASAC_ARCH}
\end{figure*}

\section{Sequence Aware SAC (SA-SAC)}

In this section, first, we discuss the chosen states, actions, and rewards, then how GRU and Decision Transformer (DT) architectures are integrated into both the actor and critic networks to formulate the sequence-aware SAC (SA-SAC) as shown in Figure \ref{fig:SASAC_ARCH}.

\subsection{State, Action, Reward}
The following are selected as states: 

\begin{itemize}
    \item \textbf{Battery SOC}: To know how much power to request from the engine for maintaining the desired SOC.
    \item \textbf{Distance Travelled}: To track how much of the cycle is completed for better planning. This information can be obtained in trucks via GPS \cite{rolando2024SAC}. 
    \item \textbf{Power Required from EM}: To know the power demand at a given time. Since the engine is decoupled from the wheels, the velocity, acceleration, and grade can be represented through the power requirement calculation from the EM.
\end{itemize}

Thus, the state \(s_t\) at time \(t\) can be represented as:
\[
s_t = \left[\text{SOC}_t, \text{D}_t, \text{P}_{\text{EM}, t}\right]
\]

Since this is a series HEV, the engine is decoupled from the wheels, meaning the engine speed is independent of the wheel speed. Using engine speed and torque as action outputs allows the SAC agent to learn the best possible engine operating points for minimal fuel consumption that meet the power demands. 

\[
a_t = \left[ \omega_{\text{eng}, t}, T_{\text{eng}, t} \right]
\]

The goal of the reward is to reduce fuel consumption while maintaining the battery's State of Charge (SOC) within the 85\% and 15\% limits throughout the day of operation. For more generalization, during training, the initial SOC was randomly chosen between [85\%, 75\%, 65\%, 55\%, 45\%] for each episode. The fuel penalty, as seen in Equation \ref{eq:main_reward}, was selected so that a higher initial SOC would require less fuel consumption compared to a lower SOC. 

\begin{equation}
\label{eq:main_reward}
R_{\text{main}} = - w_{\text{fuel}} \cdot \text{Fuel Consumption} \cdot SOC_{initial}^2 - R_{SOC}
\end{equation}
where \( w_{\text{fuel}}=5 \).

Instead of having constraints on battery SOC to be within 0 and 100\%, we use reward shaping to guide the agent for the final SOC. In Equation \ref{eq:SOC_reward}, the reward is highest if SOC stays between 18\% and 15\%, and heavy penalties if the SOC falls below 15\% or above 85\%, regardless of initial SOC.

\begin{equation}
\label{eq:SOC_reward}
R_{\text{SOC}} = 
\begin{cases} 
- w_{\text{SOC}, low} \cdot (\text{SOC} - 15\%) & \text{if SOC} < 15\% \\
 w_{\text{SOC}, good} \cdot (\text{SOC} - 15\%) & \text{if SOC} \in [15\%, 18\%] \\
0 & \text{if SOC} \in [18\%, 85\%] \\
 w_{\text{SOC}, high} \cdot (85\% - \text{SOC}) & \text{if SOC} > 85\%
\end{cases}
\end{equation}
where \( w_{\text{SOC,low}}=15 \), \( w_{\text{SOC,good}}=2.5 \) and \( w_{\text{SOC,high}}=10 \).

Finally, the states, actions, and rewards are normalized between -1 and 1 for the SAC agent to learn them easily. 

\subsection{Formulating SA-SAC}

The standard SAC code is based on the cleanRL implementation of continuous SAC \cite{cleanRLSAC}. The following changes were made to it:
\begin{itemize}
    \item \textbf{Network Architectures:}
  \begin{itemize}
      \item \textbf{GRU Actor:} GRU networks are implemented using the built-in PyTorch module. The hidden states, used for action selection, are reset at the start of each episode.
     \item \textbf{GRU Critic:} Given the input sequence of states and actions, the critic will output a sequential Q-value with each corresponding to their input timestep.  
      \item \textbf{DT Actor:} The original DT implementation from the authors’ GitHub \cite{DT_Github} was used, with two additional FFNs, after predicting the next action to output the mean and log standard deviation.
      \item \textbf{DT Critic:} The forward function was modified to predict the next return as the Q-value, given the state, action, and return trajectories. A higher future return means better performance.
  \end{itemize}

    \item \textbf{Inference Input Padding:} During inference, if the available data is shorter than the needed sequence length for GRU, DT inputs, the beginning of the sequences is padded with initial values.

    \item \textbf{Sequential Replay Buffer:} Unlike standard SAC, which samples individual transitions, this replay buffer stores trajectories for GRU and DT. During training, sequences of length $k$ are sampled to maintain temporal relationships while the initial timestep is chosen at random.

    \item \textbf{Training:} 
    The final timestep values are used from the critic, and log pi to keep the standard SAC loss calculations for policy and entropy, while the critic losses are changed as follows:
    \begin{itemize}
        \item \textbf{GRU:} Since the Q-values and target Q-values are sequences, the critic loss is computed as a single value by summing the mean squared error across the entire sequence. 
        
        \item \textbf{DT:} The DT critic loss is calculated as the difference between $R_{t+2}$ from the target Q-network and $R_{t+1}$ from the critic network.
    \end{itemize}
\end{itemize}

\noindent The final SA-SAC Pseudocode is shown in Algorithm \ref{alg:sac_variants}

\begin{algorithm}
\caption{Sequence Aware Soft Actor-Critic for SHEV}
\label{alg:sac_variants}
\textbf{Input}: Initial policy $\pi_\theta$, critics $Q_{\phi_1}, Q_{\phi_2}$, target critics $\bar{Q}_{\phi_1}, \bar{Q}_{\phi_2}$, replay buffer $\mathcal{D}$, batch size $N$, Entropy target $\mathcal{H}_{\text{target}}$\\
\textbf{Parameter}: Sequence length $k$, learning rate $\eta$, temperature $\alpha$, variant type $\in \{\text{FFN}, \text{GRU}, \text{DT}\}$\\
\textbf{Output}: Trained policy $\pi_\theta$
\begin{algorithmic}[1]
\STATE Initialize all networks and replay buffer $\mathcal{D}$
\FOR{each environment step}
    \STATE Observe state $s_t$
    \IF {variant is (GRU or DT) and $t<k$}
        \STATE Pad initial values to the input sequence beginning
    \ENDIF
    \IF {variant is SAC-GRU}
        \STATE Retrieve sequence $\{s_{t-k:t}\}$
        \STATE Encode hidden state $z_t = \text{GRU}(s_{t-k:t}, a_{t-k:t-1})$
        \STATE Sample action $a_t \sim \pi_\theta(\cdot|z_t)$
    \ELSIF{variant is SAC-DT}
        \STATE Construct trajectory $\tau_t = (R_t, s_{t-k:t}, a_{t-k:t-1})$
        \STATE Sample action $a_t \sim \pi_\theta(\cdot|\tau_t)$ using DT
        \STATE $R_t = R_{t-1}-r_t$
    \ELSE
        \STATE Sample action $a_t \sim \pi_\theta(\cdot|s_t)$ \hfill \textit{// SAC-FFN}
    \ENDIF
    \STATE Execute $a_t$ in environment, observe $r_t$, $s_{t+1}$
    \STATE Store $(s_t, a_t, (r_t $ or $ R_t), s_{t+1})$ into $\mathcal{D}$
\ENDFOR
\FOR{each training iteration}
    % \STATE Sample minibatch from $\mathcal{D}$
    \IF{variant uses sequences (GRU or DT)}
        \STATE Sample batch $N$ of sequence input: $\{s_{t-k:t}, a_{t-k:t}, (r_{t-k:t}$ or $R_{t-k:t}\}$ from $\mathcal{D}$
        \IF{GRU critic}
            \STATE $Q_{\phi}(s_t,a_t) = Q_{\phi_i}(s_{t-k}, a_{t-k})[t] $   \textit{// last critic value for actor and $\alpha$ loss}
            \STATE $\mathcal{L}(\phi_i) = \sum_{n=1}^{N} \left( \bar{Q}_{\phi_i}^{(n)} -  Q_{\phi_i}^{(n)} \right)^2$
        \ELSIF {DT Critic}
            \STATE $Q_{\phi}(s_t,a_t) = Q_{\phi_i}(s_{t-k}, a_{t-k}, R_{t-k}) =  R_{t+1} $
            \STATE $\mathcal{L}(\phi_i) =  \left( \bar{Q}_{\phi_i} -  Q_{\phi_i} \right)^2 = (R_{t+2}-R_{t+1})^2$
        \ENDIF    
    \ELSE
        \STATE Sample batch $N$ of single transitions $(s_t, a_t, r_t, s_{t+1})$ from $\mathcal{D}$
    \ENDIF
    % \STATE Compute target Q-value:
    % \[
    % y_t = r_t + \gamma \left( \min_i Q_{\bar{\phi}_i}(s_{t+1}, a_{t+1}) - \alpha \log \pi_\theta(a_{t+1}|s_{t+1}) \right)
    % \]
    \STATE $
    {J}_{\pi} = \mathbb{E}_{s_t \sim \mathcal{D},\ a_t \sim \pi_\theta} \left[ \alpha \log \pi_\theta(a_t|s_t) - Q_\phi(s_t, a_t) \right]
    $
    \STATE$
    \mathcal{L}_{\alpha} = \mathbb{E}_{a_t \sim \pi_\theta} \left[ -\alpha \left( \log \pi_\theta(a_t|s_t) + \mathcal{H}_{\text{target}} \right) \right]
    $
    \STATE Update each critic: $\phi_i \leftarrow \phi_i - \eta \nabla_{\phi_i} \mathcal{L}(\phi_i)$
    \STATE Update actor: $\theta \leftarrow \theta - \eta \nabla_{\theta} J_\pi$
    \STATE Update temperature $\alpha \leftarrow \alpha - \lambda_\alpha \nabla_\alpha \mathcal{L}_{\alpha}$
    \STATE Soft update targets: $\bar{Q}_{\phi_i} \leftarrow \tau Q_{\phi_i} + (1 - \tau)\bar{Q}_{\phi_i}$
\ENDFOR
\STATE \textbf{return} trained policy $\pi_\theta$
\end{algorithmic}
\end{algorithm}

\section{Experimental Setup and Discussion}
This study uses two software platforms: Python libraries for training due to their rich deep learning modules and efficient GPU support, while MATLAB/Simulink for validating with the forward SHEV simulator because of its accurate component modeling. The training was done with 10 cycles of the EPA Highway Fuel Economy Test (HFET), totaling 130 minutes of operation \cite{drivecycles}. 

\subsection{Hyperparameter Tuning}
We used commonly known SAC settings for the discount factor ($\gamma$), update frequency ($\tau$), and initial entropy coefficient ($\alpha$) \cite{rolando2024SAC}, while also keeping model-specific parameters such as learning rate, hidden layer sizes, and architecture depth the same across variants for fair comparisons. A gradient clipping value was applied for GRU to prevent its well-known issue of exploding gradients. Since each episode (unless specified otherwise) has 7,790 steps, to have better training efficiency, we selected training frequencies, which resulted in average episode durations of 35, 48, and 72 seconds for FFN, GRU, and DT, respectively. Final hyperparameter values are summarized in Table \ref{tab:hyperparams}.

\begin{table}[htb]
\centering

\resizebox{\columnwidth}{!}{%
\renewcommand{\arraystretch}{1}
\fontsize{18pt}{20pt}\selectfont
\newcolumntype{Y}{>{\raggedright\arraybackslash}X}

\begin{tabularx}{\textwidth}{|Y|Y|Y|Y|}
\hline
\textbf{} & \textbf{FFN} & \textbf{GRU} & \textbf{DT} \\
\hline
Learning Rate & \multicolumn{3}{c|}{1e-4} \\
\hline
Optimizer & \multicolumn{3}{c|}{Adam} \\
\hline
Batch Size & \multicolumn{3}{c|}{64} \\
\hline
$\gamma$ & \multicolumn{3}{c|}{0.99} \\
\hline
$\tau$ & \multicolumn{3}{c|}{0.005} \\
\hline
$\alpha$ & \multicolumn{3}{c|}{Auto (tuned)} \\
\hline
Replay Buffer & \multicolumn{3}{c|}{1M Capacity} \\
\hline
Actor/Critic Structure & [128, 128] 2 Layer & GRU(128, 2 Layers) & 128, 1 layer (4 heads) \\
\hline
Gradient Clipping & None & 0.25 & None \\
\hline
Training Frequency & 5 steps & 25 steps & 50 steps \\
\hline
\end{tabularx}
}
\caption{Hyperparameters used for SAC and model variants}
\label{tab:hyperparams}
\end{table}

\subsection{Ablation Study}
\begin{figure*}[!h]
    \centering
    \includegraphics[width=1.0\linewidth]{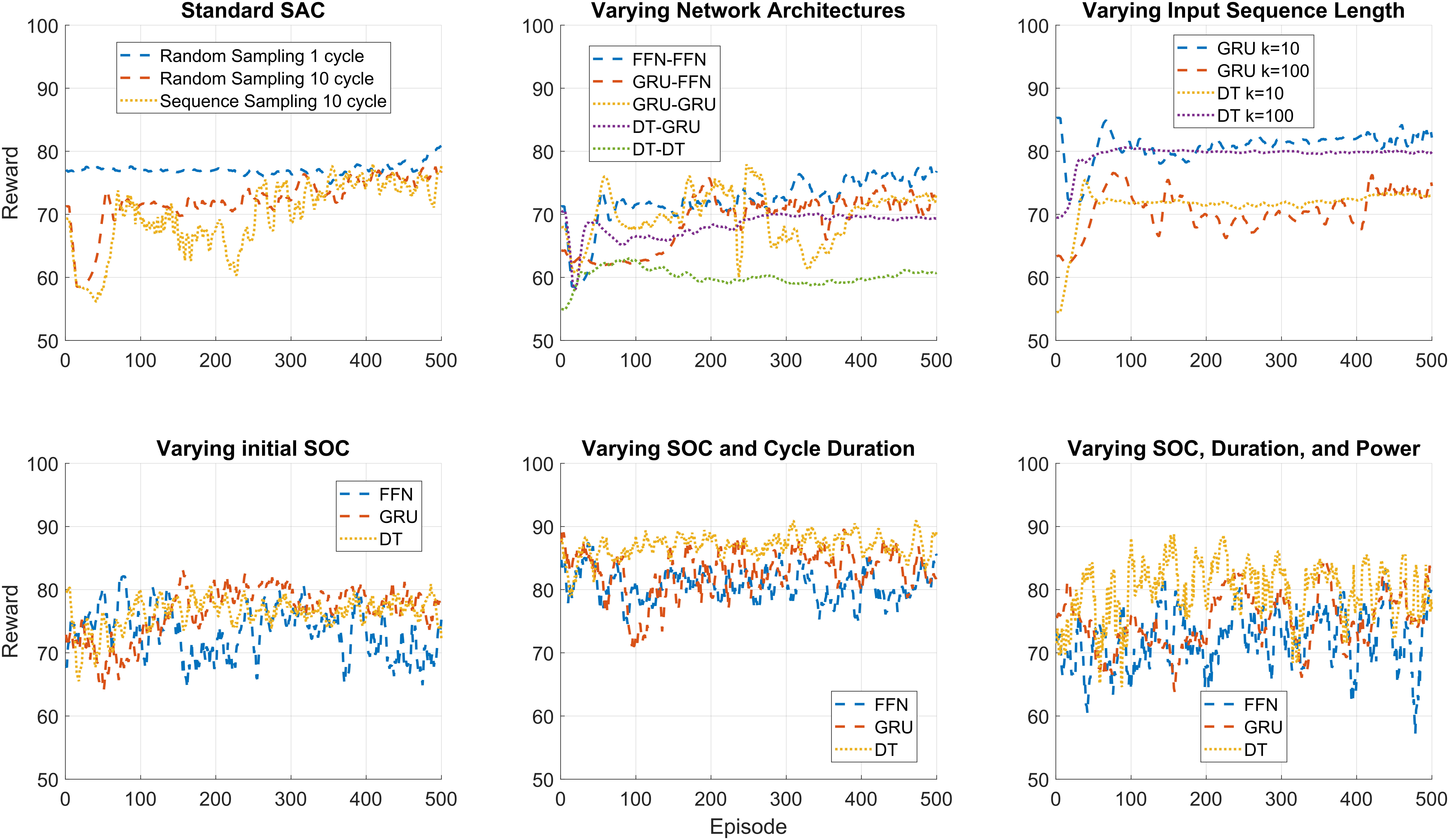}
    \caption{Ablation Study Results}
    \label{fig:Ablation Study}
\end{figure*}

The first three studies used a fixed initial SOC of 85\% to observe convergences, while the last three involved varying states. Figure \ref{fig:Ablation Study} plots the episode vs reward, where the total reward per episode is normalized by the total steps, giving values between 0 and 100 for easier comparison across conditions. A moving average with a 10-episode window is applied to the reward to reduce any noise from a single poor episode. The randomness seed was kept at one for all scenarios. The studies and their findings are as follows:

\begin{enumerate}
    \item \textbf{Sequence vs. Random Sampling for FFN (1 cycle vs. 10 cycles):} \\
    % In this study, we compare FFNs trained with random and sequential sampling. 
    FFNs are trained on 1 cycle and 10 cycles with random sampling, and then 10 cycles with sequential sampling. 
      \begin{itemize}
        \item \textbf{Findings:} SAC FFN agent converges faster with 1 cycle of random sampling but struggles with 10 cycles. Switching to sequential sampling for 10 cycles doesn't improve performance. This shows that FFNs struggle with longer episodes, and sequential sampling doesn’t help them learn better.
      \end{itemize}

    \item \textbf{Varying Actor-Critic Architecture:} \\
    This study explores different actor-critic combinations of FFN, GRU, and DT. For example, DT-GRU means a DT actor and a GRU critic. The agents were trained on ten cycles, and the input sequence was kept at 1 for a fair comparison with FFN.
      \begin{itemize}
        \item \textbf{Findings:} The Figure shows that FFN has the best improvement at one-step inputs, while DT-based actor networks converge faster and GRU-based networks show more fluctuations in performance. The DT critic performs poorly compared to the GRU critic. So, for the DT-based actor, GRU was used as the critic. FFN performs well as a critic for both GRU and FFN actors, but since FFN cannot handle sequences, GRU was used as the critic in the GRU-based actor network. 
      \end{itemize}

  \item \textbf{Varying Input Sequence Length ($k$):} \\
  In this test, input sequences are varied for $k = 10$ and $k = 100$ to find the best context length.
    \begin{itemize}
      \item \textbf{Findings:} A Longer sequence of $k=100$ improves performance for DT, as it can use more temporal information for decision-making. However, shorter sequences ($k=10$) are more effective for GRUs, which are by default well suited for capturing short-term dependencies. Both of them converged in 50 and 200 episodes, compared to FFN, which did not converge even after 500 episodes. While testing other longer sequences for DT ($k=500$ and $k=1000$), it improved performance slightly with much higher training times.
    \end{itemize}

  \item \textbf{Varying Initial SOC:} \\
  The initial State of Charge (SOC) is varied to see its impact on convergence and generalization. We continued training from the best-performing models: DT-GRU with $k=100$ and GRU-GRU with $k=10$ from the previous study.
  \begin{itemize}
    \item \textbf{Findings:} DT and GRU-based actor networks had less variance compared to the FFN-based actor. This shows their robustness to change in initial conditions.
  \end{itemize}

  \item \textbf{Varying SOC and Cycle Duration:} \\
  In addition to the initial SOC varying, the number of driving cycles used for training was varied from 1 to 10 at random. The trained agents from the previous study was continued in training. 
  \begin{itemize}
    \item \textbf{Findings:} As seen before, the DT was consistent across episodes, while GRU initially struggled but was able to outperform FFN by the end.
  \end{itemize}

  \item \textbf{Varying SOC, Duration, and Power:} \\
  This final study varies all three states (SOC, cycle duration, and power) to assess the model's robustness to simultaneous changes. The original power required from EM was multiplied by a uniformly random number between 0.5 and 1.5 for the total duration. 
  \begin{itemize}
    \item \textbf{Findings:} DT had variance at the higher end of the rewards, GRU in the middle, and FFN had the variance in the lower end of the rewards. This shows that the sequence-aware agents can generalize well with these variations.
  \end{itemize}

\end{enumerate}

In summary, the ablation study results show that SAC agents with GRU and DT networks are better at handling long episodes, fast convergences for non-varying states, and are more robust to varying state conditions compared to the FFN agents.

\subsection{Validation}

The weights of the highest-performing agent from each architectural variation were selected and tested in the Simulink SHEV forward simulator. For comparison, the Dynamic Programming (DP) baseline was solved using the DynaProg toolbox in MATLAB \cite{miretti2021dynaprog}. For the testing phase, the initial SOC was kept at 85\% for DP and SAC agents, while the final condition for DP was specified to be between 15\% and 18\% of SOC. The SAC agents are evaluated based on how close the fuel economy and final battery SOC are to DP solutions.

In addition to the HFET training cycle, two unseen highway cycles were selected: US06 Highway and Heavy Heavy-Duty Diesel Truck (HHDDT) cruise. The US06 cycle \cite{drivecycles}, representing aggressive driving, has a peak power requirement of 394 kW, compared to HFET’s 217 kW. To match HFET's 130-minute total duration, the 6-minute US06 cycle was repeated 21 times. The HHHDT cycle, with similar power requirements to HFET but longer segments at higher speeds, lasts 34 minutes and was repeated five times to achieve a total duration of 173 minutes. These two test cases assess the SAC agents' robustness under higher power demands and longer durations.

\begin{figure}[htb]
    \centering
    \includegraphics[width=\linewidth]{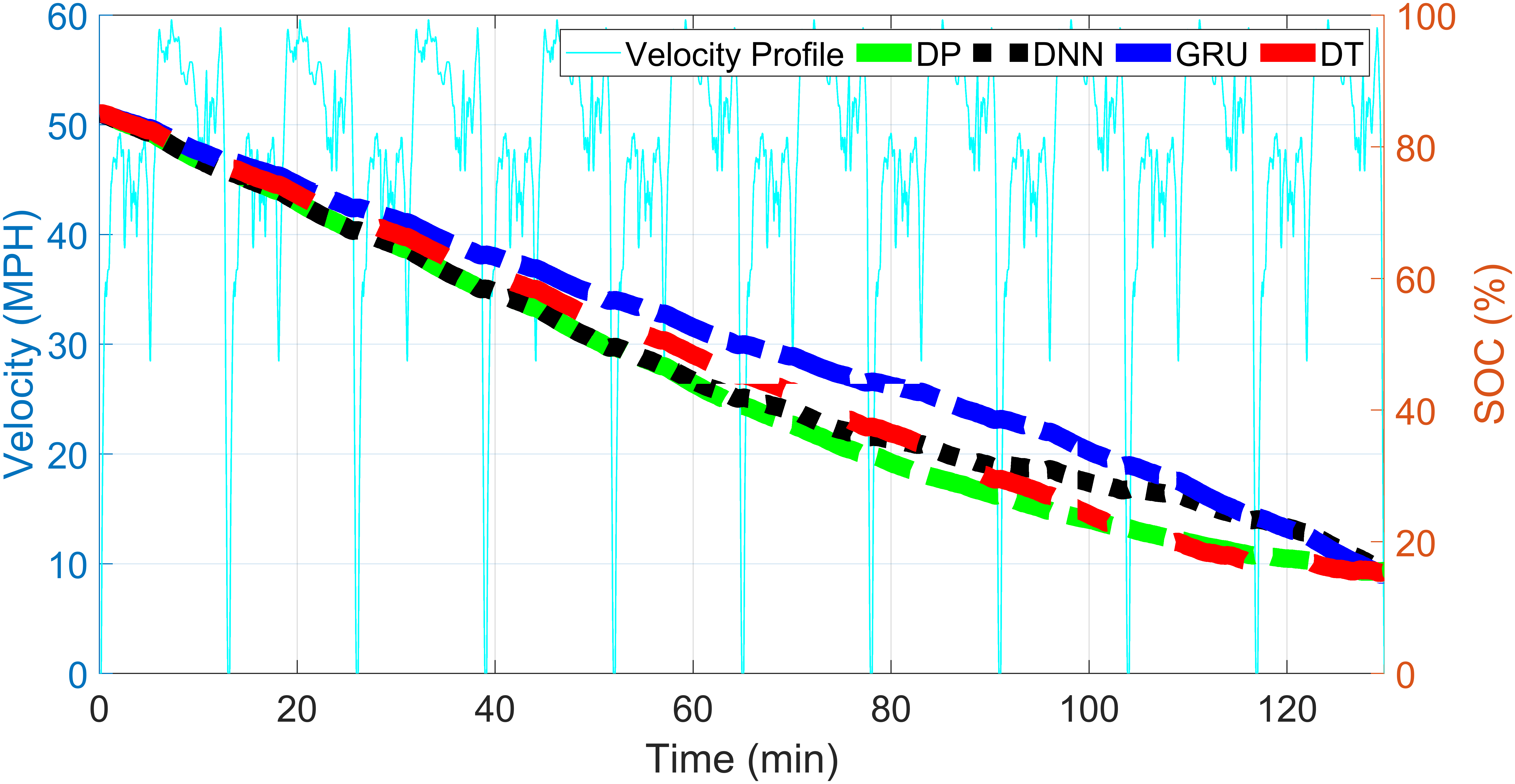}
    \caption{HFET Velocity Profile with SOC trajectories}
    \label{fig:HFET-10 SOC}
\end{figure}
\begin{figure}[htb]
    \centering
    \includegraphics[width=\linewidth]{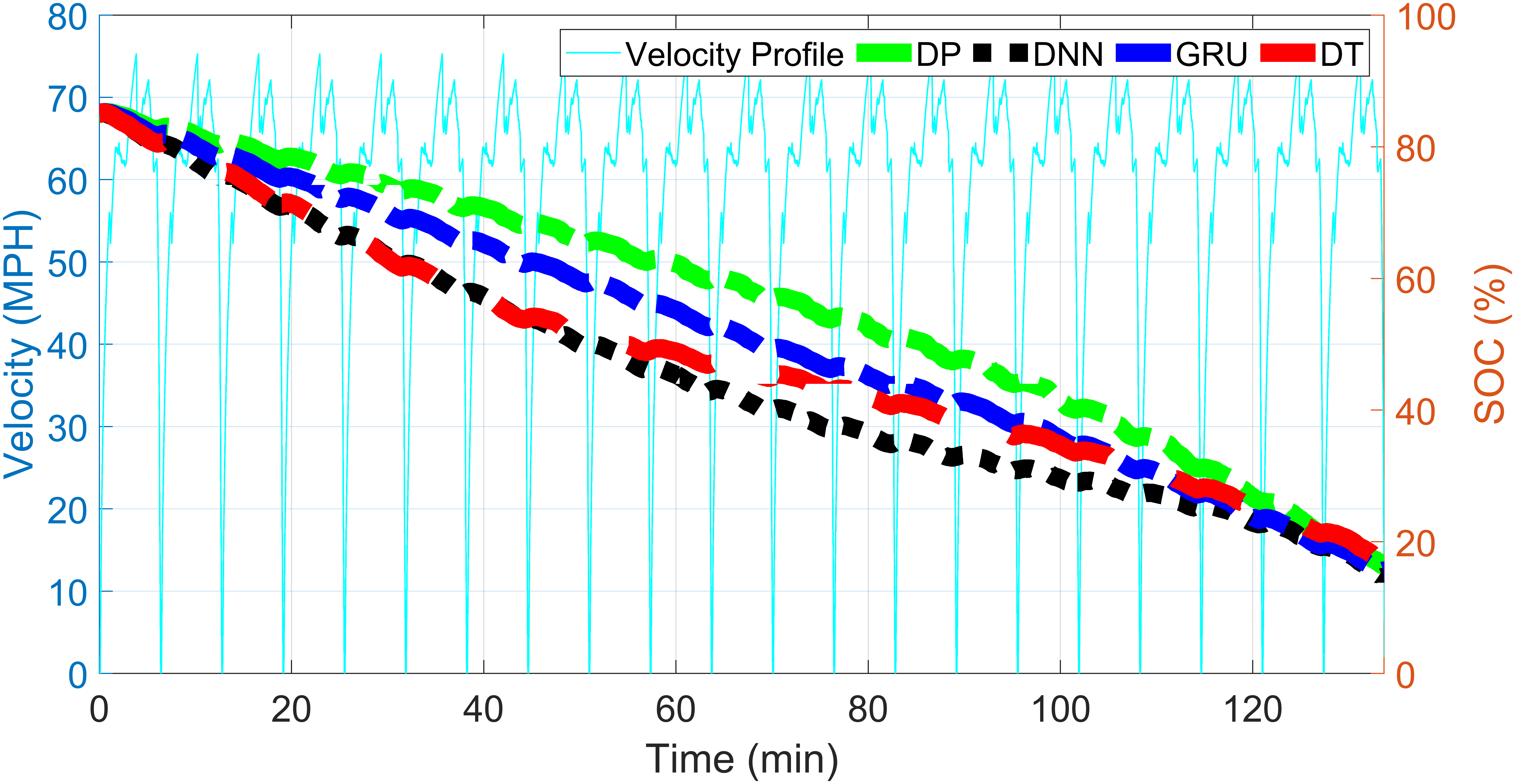}
    \caption{US06 Velocity Profile with SOC trajectories}
    \label{fig:US06 SOC}
\end{figure}
\begin{figure}[htb]
    \centering
    \includegraphics[width=1\linewidth]{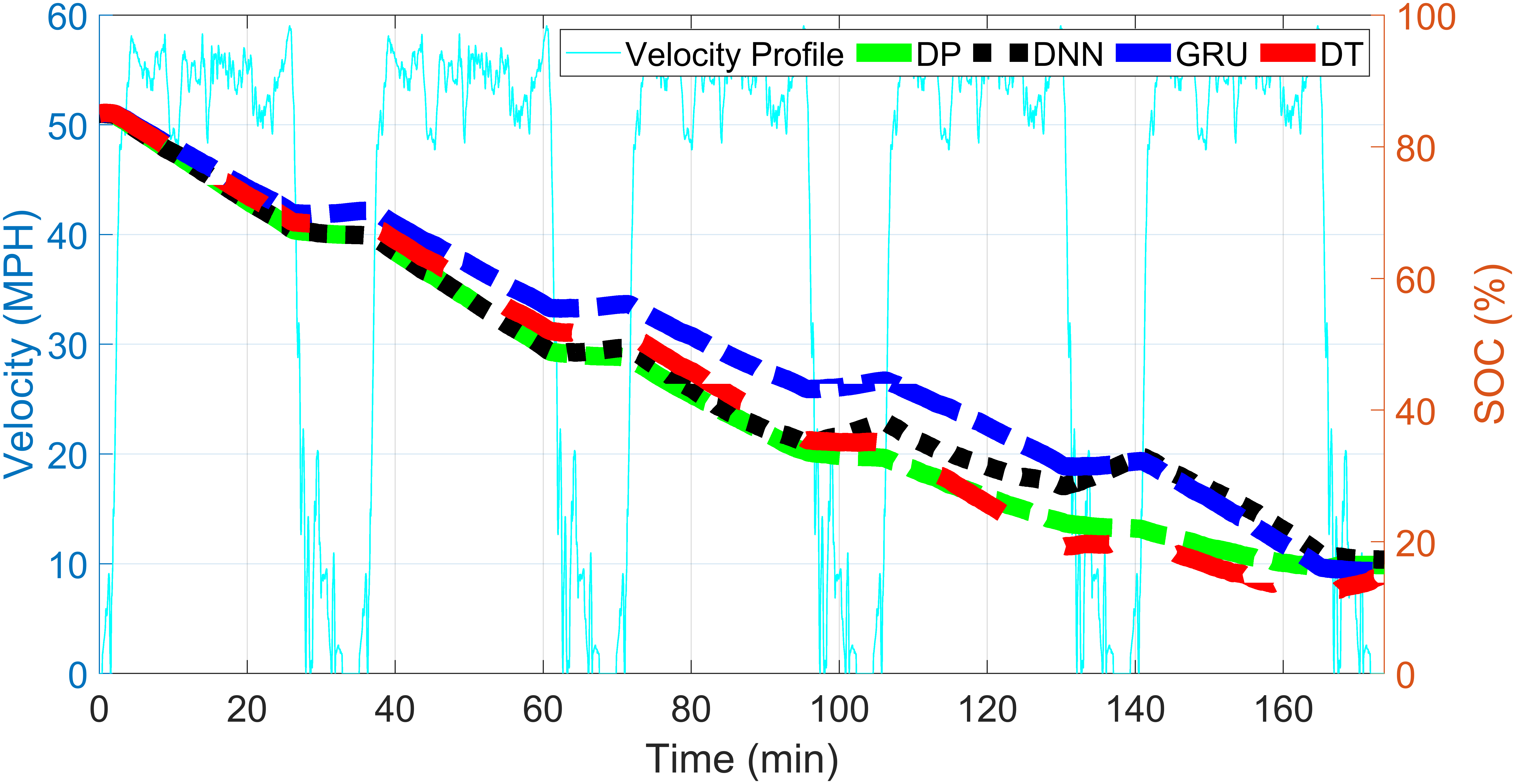}
    \caption{HHDDT Cruise Velocity Profile with SOC trajectories}
    \label{fig:HHDDT SOC}
\end{figure}
\begin{figure}[htb]
    \centering
    \includegraphics[width=1\linewidth]{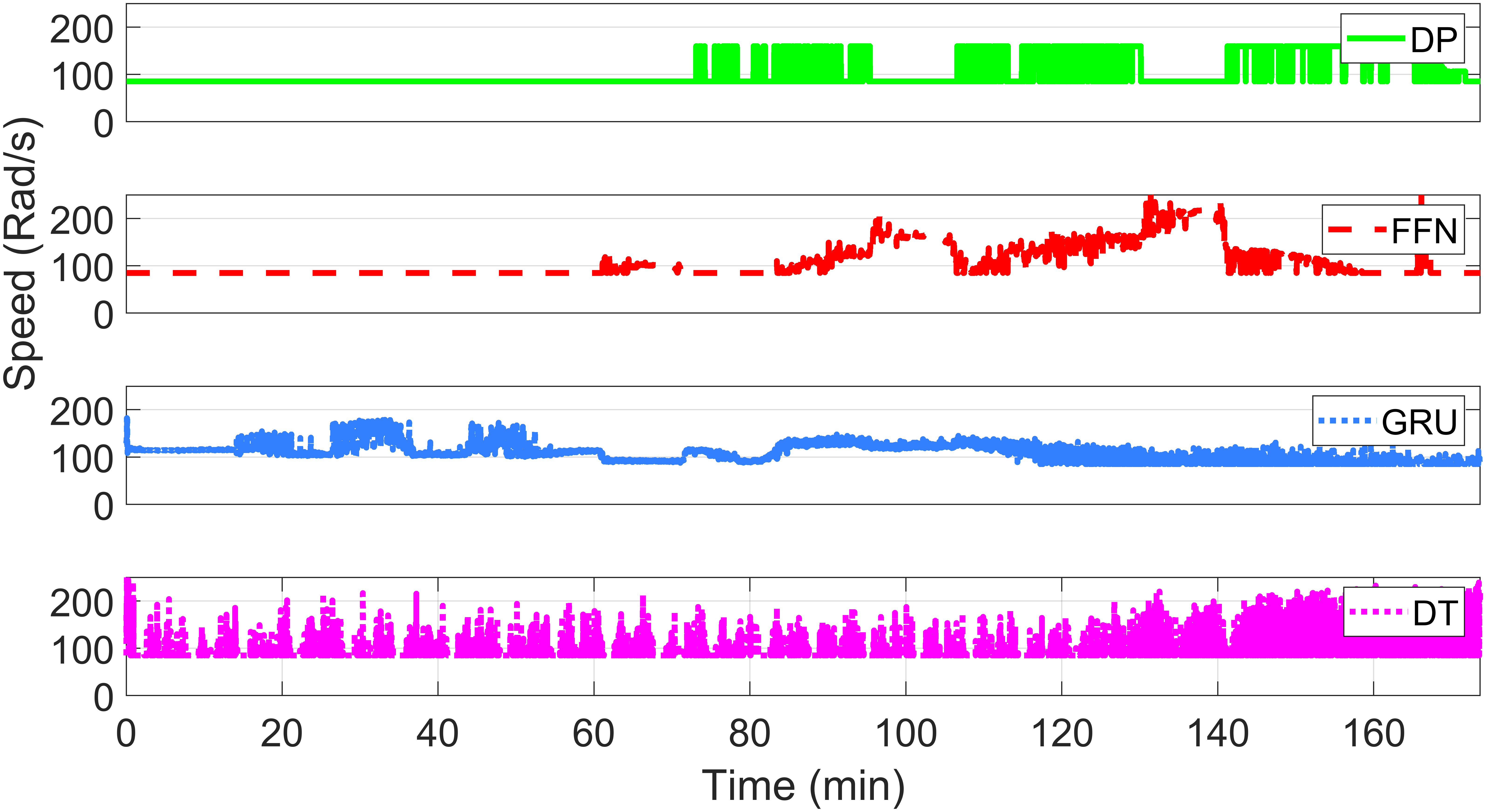}
    \caption{HHDDT Cruise Engine Speed Profile}
    \label{fig:HHDDT Speed}
\end{figure}
\begin{figure}[htb]
    \centering
    \includegraphics[width=1\linewidth]{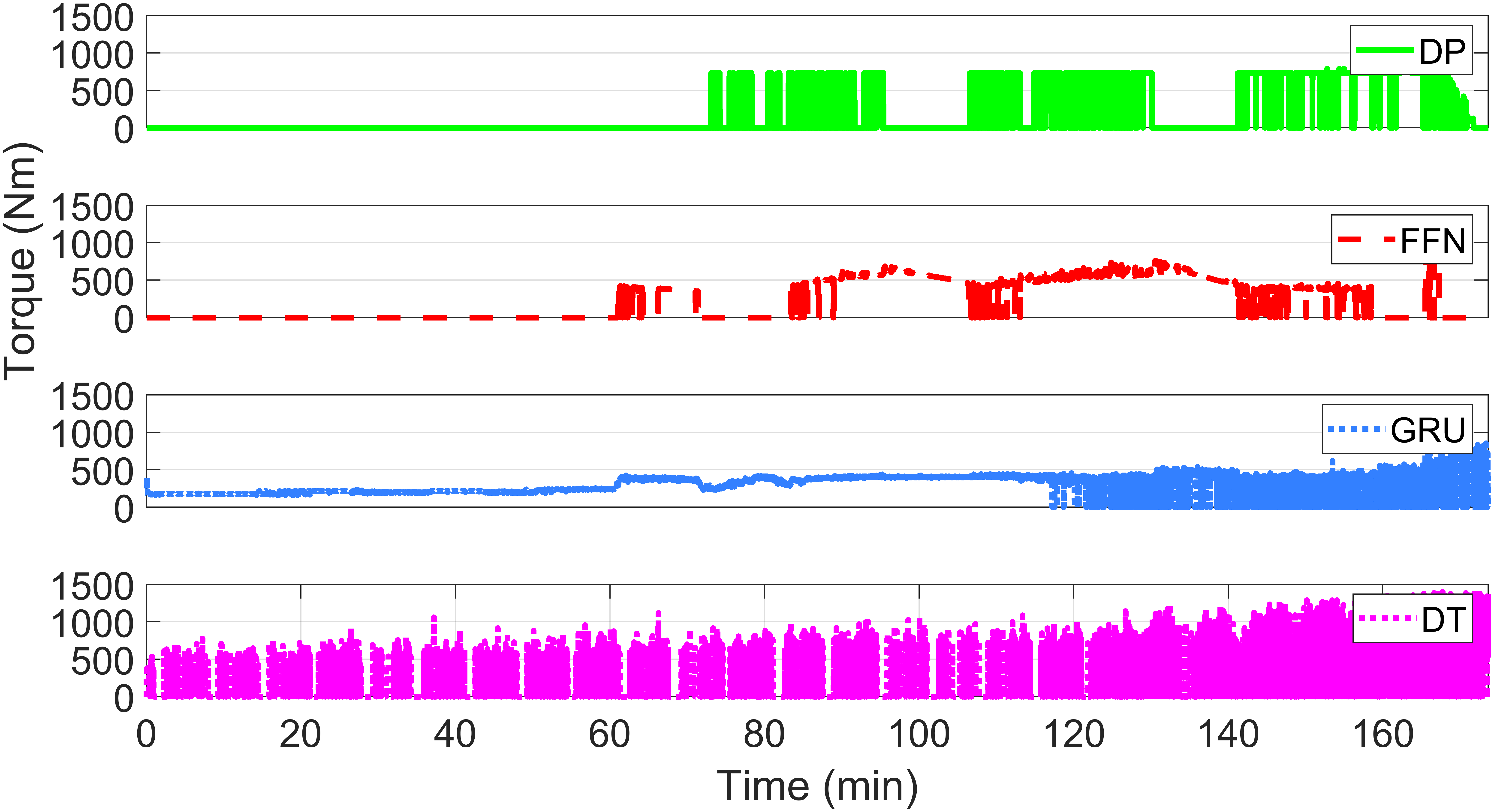}
    \caption{HHDDT Cruise Engine Torque Profile}
    \label{fig:HHDDT torq}
\end{figure}

Initially, agents with fixed SOC, cycle, and power (trained on the HFET ten cycles from the ablation study three) were tested on these three test cycles. While DT, GRU, and FFN performed within 1.8\%, 3.16\%, and 3.43\% of the DP baseline for HFET, they struggled on the unseen US06 and HHDDT cycles with more than 20\% performance degradations. To improve generalization, agents trained with varying states from the ablation study six are tested, and the results presented here are from those tests. \\
\indent The three plots (Figures \ref{fig:HFET-10 SOC}, \ref{fig:US06 SOC}, and \ref{fig:HHDDT SOC}) show how each algorithms, DP, GRU, DT, and FFN, manage SOC across various driving cycles. GRU, a recurrent model, captures short- and medium-term temporal patterns, leading to a steady, linear SOC depletion. In contrast, DT uses battery power more at first before gradually charging, similar to FFN’s approach. For the HHDDT cycle, Figures \ref{fig:HHDDT Speed} and \ref{fig:HHDDT torq} compare the engine speed and torque outputs from the SAC agents. While FFN initially idles the engine, similar to DP, GRU maintains a consistent low power output. DT, however, shows more fluctuations in the output, gradually increasing power as the cycle progresses. This noisy behavior from DT is suboptimal for actual engine performance, as the slow response and fluctuations are not ideal in real-world applications.

% Requires: \usepackage{graphicx}
\begin{table}[htb]
    \centering
    \resizebox{\columnwidth}{!}{%
    \begin{tabular}{|c|c|c|c|c|c|c|}
    \hline
    & \multicolumn{2}{c|}{HFET} & \multicolumn{2}{c|}{US06} & \multicolumn{2}{c|}{HHDDT} \\
    & SoC\textsubscript{f} (\%)& MPG & SoC\textsubscript{f} (\%)& MPG & SoC\textsubscript{f} (\%)& MPG \\ 
    \hline
    DP & 15.55& 23.71& 16.44& 4.63& 16.45& 21.83\\ \hline
    FFN & 15.81& 20.73& 14.67& 4.27& 17.29& 18.82\\
    $\Delta$ (\%)& +1.68& -12.57& -10.73& -7.72& +5.11& -13.81\\
 Total (\%)& \multicolumn{2}{|c|}{-10.89}& \multicolumn{2}{|c|}{-18.45}& \multicolumn{2}{|c|}{\textbf{-8.70}}\\ 
 \hline
    GRU & 15.10& 21.07& 15.63& 4.43& 15.59& 19.04\\ 
    $\Delta$ (\%)& -2.93& -11.14& -4.92& -4.24& -5.24& -12.8\\
 Total (\%)& \multicolumn{2}{|c|}{-14.07}& \multicolumn{2}{|c|}{-9.16}& \multicolumn{2}{|c|}{-18.04}\\ 
 \hline
    DT & 15.38& 21.68& 17.58& 4.042& 15.23& 20.75\\ 
    $\Delta$ (\%)& -1.1& -8.54& +6.93& -12.69& -7.44& -4.93\\ 
    % \hline
 Total (\%)& \multicolumn{2}{|c|}{\textbf{-9.64}}& \multicolumn{2}{|c|}{\textbf{-5.76}}& \multicolumn{2}{|c|}{-12.37}\\
 \hline
    \end{tabular}
    }
    \caption{Comparison of SoC\textsubscript{f} and Fuel Economy between DP and SAC agents: FFN, GRU, DT}
    \label{tab:comparison}
\end{table}

Table \ref{tab:comparison} shows the final SOC, MPG, and their differences compared to DP across all three networks. Since the goal was to reduce fuel consumption while maintaining a final SOC between 15\% and 18\%, the total differences of both were considered for a better representation. In all three cycles, either GRU or DT had an MPG closer to DP than FFN. While GRU struggled to maintain the final SOC near DP, it consistently stayed within the acceptable range of above 15\%. In the high power-demand US06 cycle, FFN failed to maintain the required SOC. In contrast, DT successfully balanced both MPG and SOC, staying closer to DP. The noisy outputs of DT are a result of its strategy to find the optimal engine speed and torque combination, idling when it cannot achieve this. Although this approach requires further study and adaptation to reduce noise, it still leads to better overall performance.

In conclusion, sequence-aware SAC agents perform better in both fixed and varying state conditions compared to FFN, achieving results closer to DP. However, further fine-tuning is necessary to minimize the noisy outputs and improve the model’s real-world applicability.

\section{Conclusion}

This paper explores the use of sequence-aware network architectures, GRU and DT, within SAC for SHEV engine control. The ablation study highlights that these architectures outperform FFN in generalization and convergence, as they are better at capturing temporal dependencies and adapting to dynamic driving conditions. GRU excels in handling short- and medium-term dependencies, while DT, with its ability to model long-term sequences, provides a robust solution for energy management in complex cycles. Although these networks require more computational resources, their superior performance in generalization justifies their potential over FFN.

A key challenge is the inference time for DT in the MATLAB/Simulink environment, which needs optimization for real-time use. Additionally, DT’s output was noisy and requires further fine-tuning. Since heavy-duty trucks mainly operate on highways, this study focused on highway cycles, but future work will include more compact cycles combining urban and highway driving, training the model for over 10 hours to better reflect real-world conditions. To speed up convergence, we plan to use Prioritized Experience Replay (PER) or Sequential PER in the future, which will focus training on states with higher errors, improving efficiency and performance.

\bibliography{BIB}

\end{document}